\documentclass[aps,pra,showpacs,amsmath,amssymb,amsfonts,lengthcheck,superscriptaddress,floatfix]{revtex4-2}

\usepackage{graphicx}\graphicspath{{figures/}}
\usepackage[caption=false]{subfig}
\usepackage[colorlinks=true,allcolors=blue]{hyperref}

\newcommand{\pd}{\partial}
\newcommand{\bra}[1]{\langle #1|}
\newcommand{\ket}[1]{|#1\rangle}

\newcommand{\bla}{\noindent bla\\bla\\bla\\bla\\bla}
\newcommand{\field}{B}

\begin{document}

\title{Assessing the performance of quantum annealing with nonlinear driving}

\author{Artur Soriani}
\email{asorianialves@gmail.com}
\affiliation{Instituto de F\'isica `Gleb Wataghin', Universidade Estadual de Campinas, 13083-859, Campinas, S\~{a}o Paulo, Brazil}
\author{Pierre Naz\'e}
\email{p.naze@ifi.unicamp.br}
\affiliation{Instituto de F\'isica `Gleb Wataghin', Universidade Estadual de Campinas, 13083-859, Campinas, S\~{a}o Paulo, Brazil}
\author{Marcus V. S. Bonan\c{c}a}
\affiliation{Instituto de F\'isica `Gleb Wataghin', Universidade Estadual de Campinas, 13083-859, Campinas, S\~{a}o Paulo, Brazil}
\author{Bart{\l}omiej Gardas}
\affiliation{Institute of Theoretical and Applied Informatics, Polish Academy of Sciences, Ba{\l}tycka 5, 44-100 Gliwice, Poland}
\author{Sebastian Deffner}
\affiliation{Department of Physics, University of Maryland, Baltimore County, Baltimore, Maryland 21250, USA}
\affiliation{Instituto de F\'isica `Gleb Wataghin', Universidade Estadual de Campinas, 13083-859, Campinas, S\~{a}o Paulo, Brazil}

\date{\today}

\begin{abstract}
Current generation quantum annealers have already proven to be successful problem-solvers.
Yet, quantum annealing is still very much in its infancy, with suboptimal applicability.
For instance, to date it is still an open question which annealing protocol causes the fewest diabatic excitations for a given eigenspectrum, and even whether there is a universally optimal strategy.
Therefore, in this paper, we report analytical and numerical studies of the diabatic excitations arising from nonlinear protocols applied to the transverse field Ising chain, the exactly solvable model that serves as a quantum annealing playground.
Our analysis focuses on several driving schemes that inhibit or facilitate the dynamic phases discussed in a previous work.
Rather remarkably, we find that the paradigmatic Kibble-Zurek behavior can be suppressed with ``pauses'' in the evolution, both for crossing and for stopping at the quantum critical point of the system.
\end{abstract}

\maketitle

\section{\label{sec:Intro}Introduction}

Much has been written about the anticipated \emph{quantum advantage} that novel computing technologies might demonstrate~\cite{Becher22}.
Yet actual demonstrations of a quantum system outperforming all existing classical alternatives are still very scarce and are typically based on highly constructed instances with little to no practical applications~\cite{king.carrasquilla.18,harris.sato.18,Arute.19,Yarkoni22}.

In particular, quantum annealing~\cite{Brady21}, closely related to adiabatic quantum computing~\cite{Albash18}, has achieved some prominence~\cite{Kadowaki1998PRE,GardasRBM18,grant2020adiabatic,domino2021trains}.
However, from a practical and algorithmic point of view, realizing fault-tolerant adiabatic quantum computing might even be more involved than other computational paradigms.
The reason originates in the fact that quantum annealers are subject to two fundamentally different sources of computational errors~\cite{Young2013,Sarovar2013}: (i) environmental noise, for which powerful error correction algorithms exist~\cite{Pudenz2015,Pastawski2016,Vinci2018}, and (ii) diabatic excitations, which are an inevitable consequence of finite-time driving.
To address these diabatic excitations, novel methods of so-called shortcuts to adiabaticity~\cite{Guery2019,Takahashi2017,Chen2021,Kang2022} or clever control strategies, such as thermodynamic control~\cite{bonanca2020}, might eventually provide a solution.
However, for the time being, no universally applicable method has been demonstrated (see, however, hardware-specific solutions~\cite{Pudenz2015,Pastawski2016,Vinci2018}).
Thus it is instrumental to carefully assess to what extent diabatic excitations occur and to quantify their impact on computational accuracy.

In previous works, we have demonstrated that modern tools of quantum stochastic thermodynamics~\cite{Gardas2018SR,thermo19} as well as perturbative methods~\cite{deffner2015} are very powerful and ready to be utilized.
In particular, in Ref.~\cite{Soriani2022}, we characterized the finite-time dynamics of the driven transverse field Ising chain, a model of undeniable importance to quantum annealing.
We were able to identify three parameter regions of a dynamic phase diagram with intrinsically different behaviors.
For simplicity, we focused exclusively on linear driving protocols in Ref.~\cite{Soriani2022}, as many other authors have done~\cite{Zurek2005,Polkovnikov2005,Dziarmaga2005,Francuz2016,Deffner2017,Esposito2020} (exceptions can be found in Refs.~\cite{Barankov2008,Sen2008,Quan2010} for analytical studies and Refs.~\cite{Susa2021,Hedge2022} for numerical studies).

However, linear protocols are typically neither optimal nor practically ideal.
Therefore, in this paper, we extend our previous analysis to nonlinear driving.
Despite its assumed universality \cite{Chandran2012}, we find that the paradigmatic Kibble-Zurek scaling only arises if the driving close to the quantum critical point is approximately linear.
This is demonstrated for protocols that drive through the critical point, as well as for drivings that stop right at the critical point.
Hence we find further evidence that nonlinear protocols with strategically chosen pauses can be beneficial in quantum annealing~\cite{Marshall2019,Chen2020PRAppl,Passarelli2019,Slutskii2019}.

The paper is sectioned as follows.
We review the basics of the system of interest, the transverse field Ising chain, in Sec.~\ref{sec:IsingChain}.
Section~\ref{sec:LZFandKZM} concerns the applicability of the Landau-Zener formula and the Kibble-Zurek mechanism to the system, while Sec.~\ref{sec:perturbationTheories} describes the general features of the perturbative theories we use, namely, adiabatic perturbation theory and linear response theory.
The predictions of these theories are then tested in Sec.~\ref{sec:CompNum} against numerical evolutions of the system.
The paper is concluded in Sec.~\ref{sec:Conclusion}.

\section{\label{sec:IsingChain}Transverse Field Ising chain}

We begin by establishing notions and notations.
In our analysis, we study the excess work as a function of process duration in the transverse field Ising (TI) chain \cite{Pfeuty1970}, a one-dimensional chain of $N$ spins with first-neighbor interactions.
Its Hamiltonian is
\begin{equation} \label{eq:TIHamiltonian}
H(\lambda) = - \frac{1}{2} \left( J \sum_{j=1}^{N} \sigma^z_j \sigma^z_{j+1} + \field(\lambda) \sum_{j=1}^N \sigma^x_j \right),
\end{equation}
where $\sigma_j^{x,z}$ are Pauli matrices of site $j$ and
\begin{equation} \label{eq:TIExternalField}
\field(\lambda) = J + \Delta \, \lambda
\end{equation}
is the external magnetic field, while $J$ and $\Delta$ are constants.
We assume periodic boundary conditions and choose units such that $\hbar = 1$.

Diagonalization of the system is straightforward.
Following the steps of Ref.~\cite{Dziarmaga2005}, we obtain (for $N$ even)
\begin{equation} \label{eq:TIHamiltonianDiagonalized}
H(\lambda) = \sum_k \epsilon_k(\lambda) \left( \gamma_k^\dag(\lambda) \gamma_k(\lambda) - 1/2 \right),
\end{equation}
where $\gamma_k^\dag(\lambda)$ and $\gamma_k(\lambda)$ are creation and annihilation operators of free fermions with dispersion
\begin{equation} \label{eq:TIDispersion}
\epsilon_k(\lambda) = \sqrt{ \big[ \field(\lambda) - J\cos(ka) \big]^2 + J^2 \sin^2(ka)} 
\end{equation}
and $a$ is the lattice constant.
In the thermodynamic limit, the momentum $k$ is a continuous variable ranging from $-\pi/a$ to $\pi/a$, and the quantum critical point (QCP) is at $\field = J$ (or $\lambda = 0$) as the energy gap for the lowest momentum $k_0 = \pi/N a \to 0$ vanishes.

Starting the evolution of the system in its ground state, we can work only with positive momentum values.
In this case, the dynamics of the TI chain can be simplified as the dynamics of $N/2$ two-level systems \cite{Dziarmaga2005,Soriani2022} (known as Landau-Zener (LZ) systems \cite{Landau1932,Zener1932,Stuckelberg1932,Majorana1932}), one for each positive value of $k$.
Thus the time-dependent state of the system can be written as
\begin{equation} \label{eq:TIDynamicState}
\ket{\psi(t)} = \bigotimes_{k>0} \Big( u_k(t) \ket{\downarrow_k} - v_k(t) \ket{\uparrow_k} \Big),
\end{equation}
where $\ket{\uparrow_k}$ and $\ket{\downarrow_k}$ form a basis of the LZ system labeled by $k$.
Substitution of Eq.~\eqref{eq:TIDynamicState} into Schr\"{o}dinger's equation takes us to the time-dependent Bogoliubov--de Gennes equations,
\begin{equation} \label{eq:TIDifferentialEquations}
\begin{split}
i \dot{u}_k(t) & =  - \big[ \field(\lambda) - J \cos(ka) \big] u_k(t) - J \sin(ka) v_k(t), \\
i \dot{v}_k(t) & =  - J \sin(ka) u_k(t) + \big[ \field(\lambda) - J \cos(ka) \big] v_k(t).
\end{split}
\end{equation}
The parametric ground state of the system can be obtained from Eq.~\eqref{eq:TIDynamicState} with the substitutions $u_k(t) \to \cos\theta_k(\lambda)$ and $v_k(t) \to \sin\theta_k(\lambda)$,
where
\begin{equation} \label{eq:TITheta}
\theta_k(\lambda) = \frac{1}{2} \arctan \left( \frac{J \sin(ka)}{\field(\lambda) - J \cos(ka)} \right).
\end{equation}

We consider processes of finite duration $\tau = t_f - t_i$ implemented through a predetermined time dependence to $\lambda = \lambda(t)$, taking the external parameter from $\lambda_i = \lambda(t_i)$ to $\lambda_f = \lambda(t_f)$.
The excess work per spin performed during the process, also referred to as residual energy or excitation energy \cite{Francuz2016,Esposito2020}, reads
\begin{equation} \label{eq:TIExcessWork}
w_{\mathrm{ex}}(\tau) = \frac{1}{N} \sum_{k>0} 2 \epsilon_k(\lambda_f) p_k(\tau),
\end{equation}
where
\begin{equation} \label{eq:TIExcitationProbability}
p_k(\tau) = \Big| \sin\theta_k(\lambda_f) u_k(t_f) - \cos\theta_k(\lambda_f) v_k(t_f) \Big|^2
\end{equation}
is the probability of creating a pair of fermions with opposite momenta $k$ and $-k$ during the evolution.
In our analysis, we fix the initial and final points of the evolution, $\lambda_i$ and $\lambda_f$ (also called end points), while considering processes with different values of $\tau$.
Consequently, we have $\dot{\lambda}(t) \sim \tau^{-1}$, $\ddot{\lambda}(t) \sim \tau^{-2}$, and so on.

In what follows, we will analyze the $\tau$ dependence of the excess work of Eq.~\eqref{eq:TIExcessWork} with nonlinear driving for two different scenarios: crossing the critical point and stopping at the critical point.
This will complement what was presented in Ref.~\cite{Soriani2022} where most of the discussion focused on only linear driving.
In order to avoid redundancy, some technical details are omitted here, but they can be found in Ref.~\cite{Soriani2022}.

\section{Landau-Zener formula and Kibble-Zurek mechanism \label{sec:LZFandKZM}}

Motivated by the results obtained for linear driving, we separate the processes to be implemented in the finite TI chain into three broadly defined groups: sudden, intermediate, and slow processes.
To determine to what group a process belongs, the duration $\tau$ must be compared with the time scale $\left( \frac{N}{\pi} \right)^2 \frac{\Delta}{J^2}$, essentially $1$ over the minimum gap squared.
As shown in Sec.~\ref{sec:CompNum}, the intermediate regime is well described by nonperturbative theories, which are the topic of this section.

\subsection{\label{sec:LZFandKZM_cross}Crossing the critical point}

In the scenario of crossing the critical point, the $\tau$ behavior of the excitation probability can be well described by the Landau-Zener (LZ) formula,
\begin{equation} \label{eq:LZFormula}
p_k^{\mathrm{LZ}}(\tau) = \exp\left( -\pi \frac{J_k^2}{\Delta \dot{\lambda}(t_c)} \right),
\end{equation}
within a certain range of $\tau$ values.
In the previous expression, $J_k = J\sin(ka)$, and $t_c$ is obtained from $\lambda(t_c) = 0$, the time of crossing the QCP.
The LZ formula leads to two different scales for the excess work.

If $J^2/\Delta \dot{\lambda}(t_c) \gg 1$, the argument of the exponential in Eq.~\eqref{eq:LZFormula} can be expanded for $ka \ll 1$.
Then, carrying out the continuous sum in Eq. \eqref{eq:TIExcessWork} in the thermodynamic limit gives (Eq.~(37) of Ref.~\cite{Soriani2022})
\begin{equation} \label{eq:crossing_KZMExcessWork}
w_{\mathrm{ex}}^{\mathrm{KZM}}(\tau) = \frac{\Delta|\lambda_f|}{2\pi}  \sqrt{\frac{\Delta \dot{\lambda}(t_c)}{J^2}},
\end{equation}
where KZM stands for the Kibble-Zurek mechanism. This phenomenological theory of second-order phase transitions, when extended to quantum phase transitions, predicts the $\tau^{-1/2}$ behavior of the average number of excitations \cite{Zurek2005,Polkovnikov2005,Dziarmaga2005} (to which the excess work is proportional, when crossing the QCP).

Increasing the process duration $\tau$, we reach a point $\frac{J^2}{\Delta \dot{\lambda}(t_c)} \left( \frac{\pi}{N} \right)^2 \sim 1$ where the lowest-momentum term in Eq.~\eqref{eq:TIExcessWork} dwarfs every other term, since the exponential of Eq.~\eqref{eq:LZFormula} is highly peaked at $k_0$.
In this case, we get (Eq.~(39) of Ref.~\cite{Soriani2022})
\begin{equation} \label{eq:crossing_LZFExcessWork}
w_{\mathrm{ex}}^{\mathrm{LZF}}(\tau) = \frac{2 \Delta |\lambda_f|}{N} \exp\Biggl[ -\pi \left( \frac{\pi}{N} \right)^2 \frac{J^2}{\Delta \dot{\lambda}(t_c)} \Biggr],
\end{equation}
which signifies that only the lowest value of $k$ contributes to the sum in Eq.~\eqref{eq:TIExcessWork}.
Consequently, the LZF result has the same characteristic exponential decay of the LZ formula in Eq.~\eqref{eq:LZFormula}, hence its name.

A comment about the LZ formula~\eqref{eq:LZFormula} is in order.
It was first obtained for linear driving \cite{Landau1932,Zener1932,Stuckelberg1932,Majorana1932}, and consequently, its extension to nonlinear protocols relies on the assumption that most of the transitions between eigenstates happen close to the time of crossing $t_c$.
Thus we can linearly approximate $\lambda$ around the time of crossing, $\lambda(t) \approx (t - t_c) \dot{\lambda}(t_c)$.
However, if $\dot{\lambda}(t_c) \ll \tau^{-1}$, this approximation is insufficient, and we need to use a higher-order polynomial approximation, which is not analytically solvable.
Therefore Eq.~\eqref{eq:LZFormula}, and the results derived from it in this section, are valid as long as the QCP is not traversed too slowly for a given $\tau$, when compared with other points of the evolution.

\subsection{\label{sec:LZFandKZM_stop}Stopping at the critical point}

In the scenario of stopping at the critical point, we fix $t_c = t_f$, which means that the evolution ends exactly when the system reaches the critical point $\lambda(t_f) = 0$.
The $\tau$ behavior of $p_k$ in Eq.~\eqref{eq:TIExcessWork} is expressed by the ``half'' LZ formula \cite{Damski2006}, which leads to (Eq.~(A3) from Appendix A of Ref.~\cite{Soriani2022})
\begin{equation} \label{eq:stopping_KZMExcessWork}
w_\mathrm{ex}^\mathrm{KZM}(\tau) = \frac{K J}{\pi} \frac{\Delta \dot{\lambda}(t_f)}{J^2},
\end{equation}
where
\begin{equation} \label{eq:stopping_KZMConstant}
\begin{split}
& K \equiv \int_0^{\infty} x \Biggl[ 1 - \exp\left( -\frac{\pi}{4} x^2 \right) \frac{\sinh\left( \frac{\pi}{2} x^2 \right)}{\pi x^2 } \\
&\times \left| \Gamma\left( 1 + \frac{i}{4} x^2 \right) + \frac{ \exp(i \pi/4) }{2} x \Gamma\left( \frac{1}{2} + \frac{i}{4} x^2 \right)\right|^2 \Biggr] dx
\end{split}
\end{equation}
is an integral that can be computed numerically (here, $\Gamma$ denotes the gamma function).
Thus, when stopping at the QCP, KZM predicts $w_{\mathrm{ex}} \sim \tau^{-1}$, and this is \emph{not} the same scale predicted for the average number of excitations \cite{Francuz2016,Esposito2020} (which always scales as $\tau^{-1/2}$).
Equation~\eqref{eq:stopping_KZMExcessWork}, much like Eqs.~\eqref{eq:LZFormula}--\eqref{eq:crossing_LZFExcessWork}, does not apply when the derivative it contains is too small (see Sec.~\ref{sec:CompNum}).

\section{\label{sec:perturbationTheories}Perturbation theories}

In contrast to the intermediate processes, sudden and slow processes can be described by perturbation theories, two of which are outlined in this section.

\subsection{\label{sec:perturbationTheories_APT}Adiabatic perturbation theory}

For slow processes, no matter the scenario, the excitation probability is given by adiabatic perturbation theory (APT) \cite{Messiah1962,Rigolin2008}.
If $\frac{J^2}{\Delta \dot{\lambda}(t_c)} \left( \frac{\pi}{N} \right)^2 \gg 1$, and if $\dot{\lambda}(t_i)$ and $\dot{\lambda}(t_f)$ are not zero, the excess work reads (Eqs.(40) and (41) of Ref.~\cite{Soriani2022})
\begin{equation} \label{eq:APT1ExcessWork}
w_{\mathrm{ex}}^{\mathrm{APT1}}(\tau) = \sum_{k>0} \frac{\Delta^2 J_k^2}{8N} \epsilon_k(\lambda_f) \left| \frac{ \dot{\lambda}(t_f)}{\epsilon_k^3(\lambda_f)} - e^{2i\phi_k(\tau)} \frac{\dot{\lambda}(t_i)}{\epsilon_k^3(\lambda_i)} \right|^2,
\end{equation}
where
\begin{equation} \label{eq:TI_DynamicPhase}
\phi_k(\tau) = - \int_{t_i}^{t_f} \epsilon_k[ \lambda(t') ] dt'.
\end{equation}
Thus, for generic protocols, the first-order term of APT dominates and gives $w_{\mathrm{ex}} \sim \tau^{-2}$.

Alternatively, if $\dot{\lambda}(t_i) = 0 = \dot{\lambda}(t_f)$ and, simultaneously, $\ddot{\lambda}(t_i)$ and $\ddot{\lambda}(t_f)$ are not zero, the second-order term of APT is the first nonzero correction, and the excess work becomes
\begin{equation} \label{eq:APT2ExcessWork}
w_{\mathrm{ex}}^{\mathrm{APT2}}(\tau) = \sum_{k>0} \frac{\Delta^2 J_k^2}{32N} \epsilon_k(\lambda_f) \left| \frac{\ddot{\lambda}(t_f)}{\epsilon_k^4(\lambda_f)} - e^{2i\phi_k(\tau)} \frac{\ddot{\lambda}(t_i)}{\epsilon_k^4(\lambda_i)} \right|^2.
\end{equation}
In this case, APT predicts $w_{\mathrm{ex}} \sim \tau^{-4}$.

In APT, the behavior around the end points $\lambda_i$ and $\lambda_f$ is crucial; yet the expressions \eqref{eq:APT1ExcessWork} and \eqref{eq:APT2ExcessWork} remain the same in each scenario (crossing or stopping at the QCP).
Note that both expressions can be extended to the thermodynamic limit.
When crossing the QCP, this is done as usual: by replacing sums with integrals (see Eq.~(42) of Ref.~\cite{Soriani2022}).
However, when stopping at the QCP, one notes that the first term inside the absolute values of Eqs.~\eqref{eq:APT1ExcessWork} and \eqref{eq:APT2ExcessWork} for $k_0 \to 0$ dominates the whole sum as $\epsilon_{k_0}(\lambda_f) \to 0$ (see Eq.~(A5) from Appendix A of Ref.~\cite{Soriani2022}).

\subsection{\label{sec:perturbationTheories_LRT}Linear response theory}

For sudden processes, for which $J^2/\Delta \dot{\lambda}(t_c) \ll 1$, linear response theory (LRT) gives an appropriate description \cite{naze2022}.
The excess work per spin from LRT reads (see Appendix B in Ref.~\cite{Soriani2022})
\begin{equation} \label{eq:LRTExcessWork}
w_{\mathrm{ex}}^\mathrm{LRT}(\tau) = \frac{1}{2} \int_{t_i}^{t_f} \int_{t_i}^{t_f} \Psi_i(t-t') \dot{\lambda}(t) \dot{\lambda}(t') dt' dt,
\end{equation}
where (Eq.~(48) of Ref.~\cite{Soriani2022})
\begin{equation} \label{eq:RelaxationFunction}
\Psi_i(t) = \frac{1}{N} \sum_{k>0} \frac{ \Delta^2 J_k^2 }{ \epsilon^3_k(\lambda_i) } \cos[ 2 \epsilon_k(\lambda_i) t ]
\end{equation}
is the system's relaxation function per spin \cite{Naze2021} obtained from the response function 
\begin{equation}
    \Phi_i(t) = -i \bra{\psi(t_i)}[\pd_{\field}H(t_i),\pd_{\field}H(t)]\ket{\psi(t_i)}
\end{equation}
through the relation
\begin{equation}
    \Phi_i(t) = -\frac{d\Psi_i(t)}{dt},
\end{equation}
where $\pd_{\field}$ is the derivative with respect to $\field$, $[\cdot,\cdot]$ is the commutator, and $\ket{\psi(t_i)}$ is the initial ground state. The operator $\pd_{\field}H(t)$ denotes $\pd_{\field}H$ in the interaction picture at time $t$.
We remark that the expression remains the same whether we cross or stop at the QCP. Moreover, in the limit $\tau\to 0$, the excess work per spin is given by \cite{Naze2021}
\begin{equation}
    w_{\mathrm{ex}}^\mathrm{LRT}(\tau)=\frac{\Psi_i(0)\delta\lambda^2}{2},\quad \delta\lambda=\lambda(t_f)-\lambda(t_i)
\end{equation}
no matter the form of the protocol $\lambda(t)$.

We expect LRT to work in the weak-driving limit, where $\Delta/J \ll 1$.
Nevertheless, we shall see that LRT can give good predictions well outside its assumed range of validity (see also Ref.~\cite{naze2022}).

\section{\label{sec:CompNum}Analytical approximations versus numerically exact solution}

The results of the previous two sections were extensively analyzed and tested in the case of linear variation of the external field; see, for instance, our recent paper~\cite{Soriani2022}.
Motivated by this earlier work, we continue to analyze to what extent the scaling behavior manifests itself for a selection of nonlinear protocols.
We choose protocols according to the expected features of the excitation probability and implement them numerically, through the application of the standard fourth-order Runge-Kutta method to Eq.~\eqref{eq:TIDifferentialEquations}.
In all of the following plots, the natural oscillations of the perturbation theories [exemplified by the dynamical phase of Eq.~\eqref{eq:TI_DynamicPhase}] were averaged out (see Ref.~\cite{Soriani2022} for comparison).

\subsection{\label{sec:ComNum_cross}Crossing the critical point}

\begin{figure}
\includegraphics[width=\columnwidth]{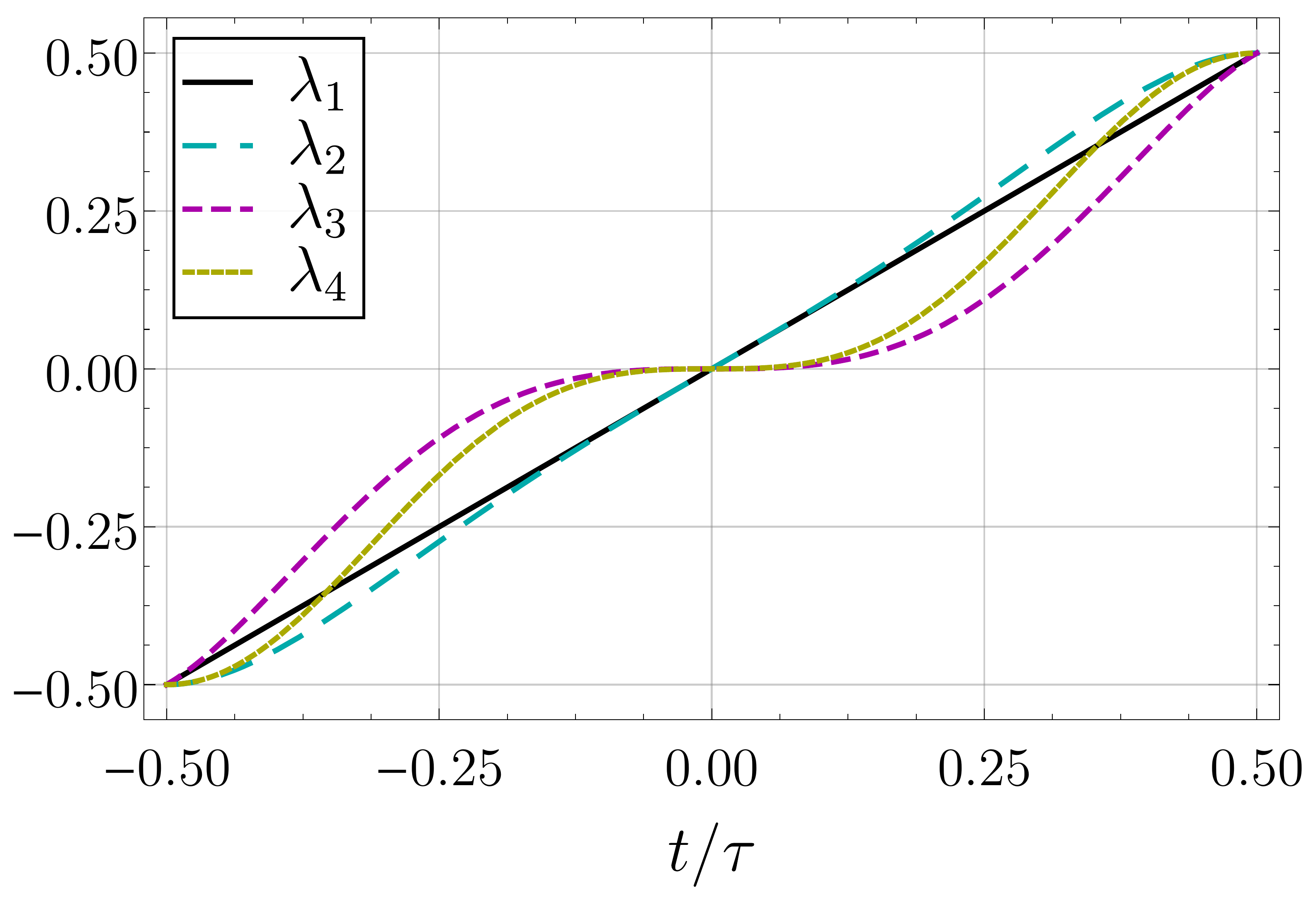}
\caption{\label{fig:crossing_protocols}
Protocols of Eqs.~\eqref{eq:crossing_lambda1}--\eqref{eq:crossing_lambda4}.
}
\end{figure}

Consider the following protocols, from $t_i = -\tau/2$ to $t_f = \tau/2$:
\begin{subequations} \label{eq:crossing_ptc}
\begin{align}
\label{eq:crossing_lambda1} \lambda_1(t) & = \frac{t}{\tau}, \\
\label{eq:crossing_lambda2} \lambda_2(t) & = -8 \left( \frac{t}{\tau} \right)^5 + 2 \left( \frac{t}{\tau} \right)^3 + \frac{t}{\tau}, \\
\label{eq:crossing_lambda3} \lambda_3(t) & = -16 \left( \frac{t}{\tau} \right)^5 + 8 \left( \frac{t}{\tau} \right)^3, \\
\label{eq:crossing_lambda4} \lambda_4(t) & = 64 \left( \frac{t}{\tau} \right)^7 - 56 \left( \frac{t}{\tau} \right)^5 + 14 \left( \frac{t}{\tau} \right)^3.
\end{align}
\end{subequations}
These polynomials are depicted in Fig.~\ref{fig:crossing_protocols}.
Note that $\lambda_1$ is the linear protocol considered previously, while the others are the simplest polynomials that fulfill the general properties $t_c = 0$, $\lambda_i = -1/2$, and $\lambda_f = 1/2$ and the following specific properties:
(i) $\lambda_2$ has $\dot{\lambda}_2(t_c) = \tau^{-1}$ and $\dot{\lambda}_2(t_i) = 0 = \dot{\lambda}_2(t_f)$;
(ii) $\lambda_3$ has $\dot{\lambda}_3(t_c) = 0$ and $\dot{\lambda}_3(t_i) = \tau^{-1} = \dot{\lambda}_3(t_f)$;
and (iii) $\lambda_4$ has $\dot{\lambda}_4(t_c) = 0$, $\dot{\lambda}_4(t_i) = 0 = \dot{\lambda}_4(t_f)$, and $\ddot{\lambda}_4(t_i) = 14 \tau^{-2} = - \ddot{\lambda}_4(t_f)$.

Protocol $\lambda_2$ was motivated by its expected power-law decay, given by Eq.~\eqref{eq:APT2ExcessWork} instead of Eq.~\eqref{eq:APT1ExcessWork}.
Conversely, $\lambda_3$ was motivated by its expected ability to escape the description of the LZ formula, Eq.~\eqref{eq:LZFormula}.
Lastly, $\lambda_4$ combines the previous two motivations.

In the following paragraphs, we will compare individually each of the nonlinear protocols [Eqs.~\eqref{eq:crossing_lambda2}-\eqref{eq:crossing_lambda4}] with the linear one [Eq.~\eqref{eq:crossing_lambda1}], highlighting their effects on the excess work.

\begin{figure*}

\subfloat[\label{fig:crossing_lambda12_APT}]{\includegraphics[width=.5\textwidth]{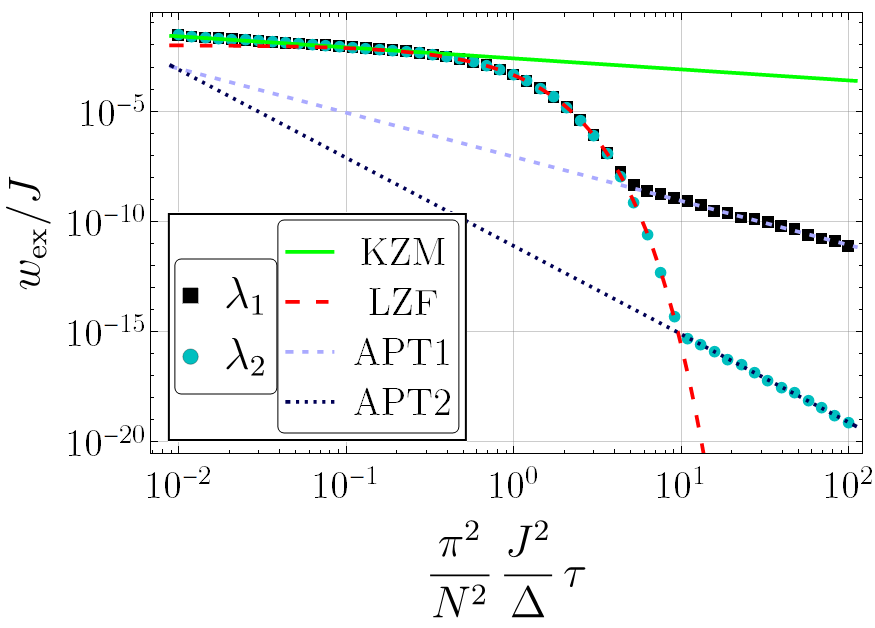}}
\subfloat[\label{fig:crossing_lambda13_APT}]{\includegraphics[width=.5\textwidth]{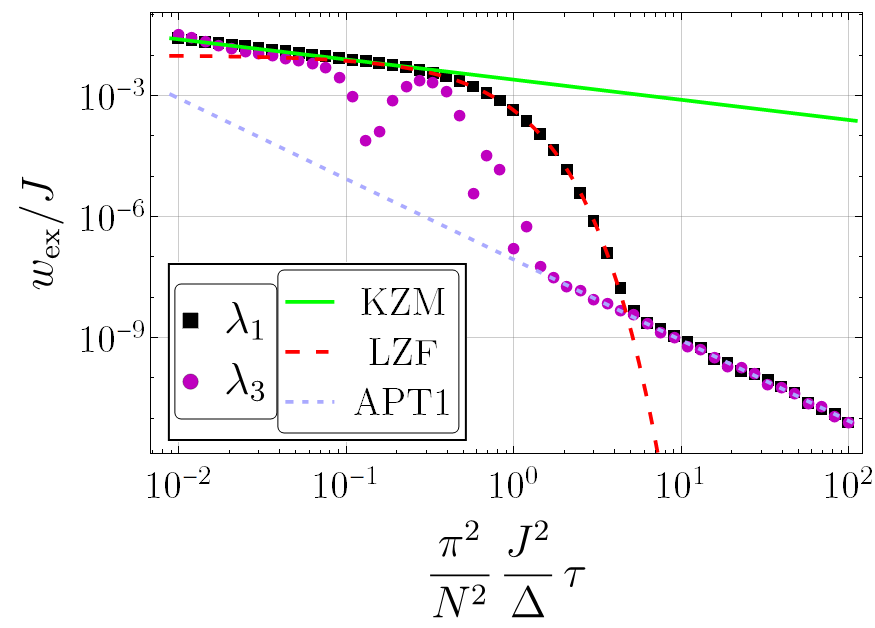}}

\subfloat[\label{fig:crossing_lambda14_APT}]{\includegraphics[width=.5\textwidth]{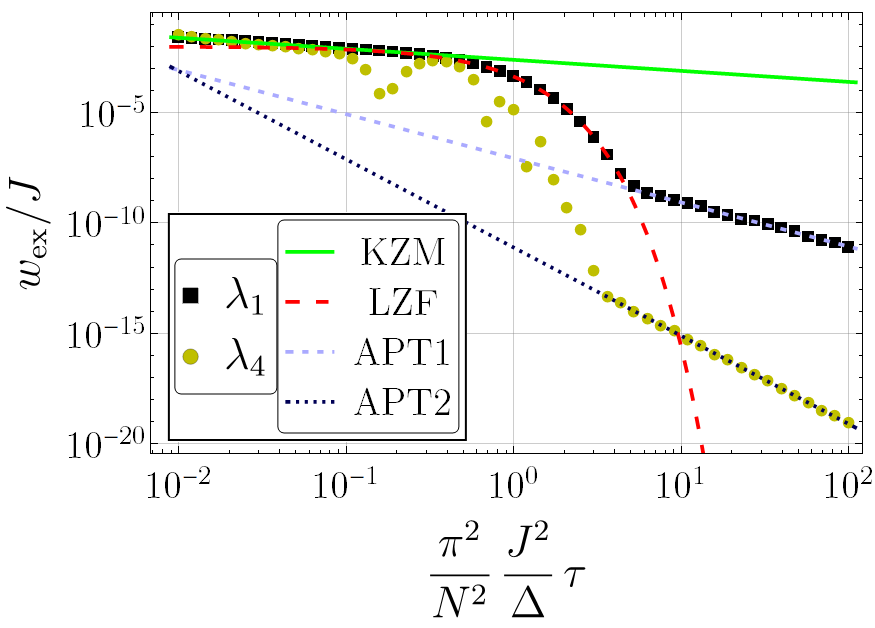}}
\subfloat[\label{fig:crossing_KZMregion}]{\includegraphics[width=.5\textwidth]{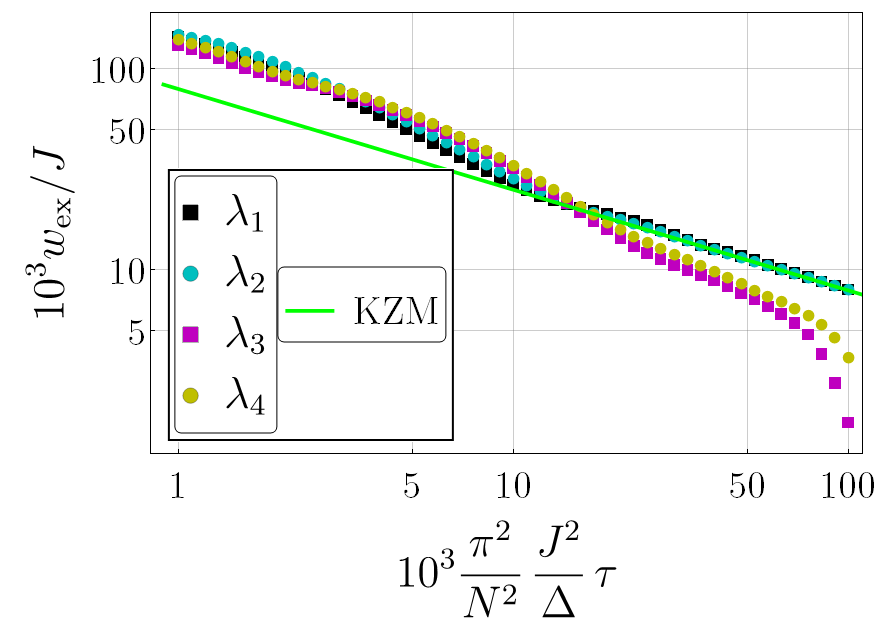}}

\caption{\label{fig:crossing_ExcessWorkPlots_APT}
Plots of the excess work for the protocols in Eqs.~\eqref{eq:crossing_lambda1}--\eqref{eq:crossing_lambda4} as a function of process duration for $\Delta/J = 1$ and $N=100$.
The symbols represent the numerics for each protocol; the green line represents KZM of Eq.~\eqref{eq:crossing_KZMExcessWork}; the red, long-dashed line represents LZF of Eq.~\eqref{eq:crossing_LZFExcessWork}; the light blue, medium-dashed line represents first-order APT of Eq.~\eqref{eq:APT1ExcessWork}; and the dark blue, short-dashed line represents second-order APT of Eq.~\eqref{eq:APT2ExcessWork}.
(a) Protocols $\lambda_1$ and $\lambda_2$ present the same crossovers discussed in Ref.~\cite{Soriani2022}, each with a different power-law scaling for slow processes.
(b) Protocols $\lambda_1$ and $\lambda_3$ display the same APT scale, but $\lambda_3$ does not agree with the KZM and LZF results.
(c) Protocols $\lambda_1$ and $\lambda_4$ differ the most, as they do not share behavior in any region of the plot.
(d) Protocols $\lambda_1$ and $\lambda_2$ have KZM and LZF scales, while $\lambda_3$ and $\lambda_4$ do not.
}

\end{figure*}

The protocol $\lambda_2$ \eqref{eq:crossing_lambda2} has the same first derivative at the QCP as the linear protocol, which means that both protocols display the KZM and LZF scales of Eqs.~\eqref{eq:crossing_KZMExcessWork} and \eqref{eq:crossing_LZFExcessWork}, respectively.
On the other hand, $\lambda_2$ differs from $\lambda_1$ in its first derivatives at the end points, which are null.
Therefore, while APT for $\lambda_1$ gives a $\tau^{-2}$ scale [given by Eq.~\eqref{eq:APT1ExcessWork}], for $\lambda_2$ it gives a $\tau^{-4}$ scale [given by Eq.~\eqref{eq:APT2ExcessWork}].
This can be seen in Fig.~\ref{fig:crossing_lambda12_APT}.
Note that, while the APT result for $\lambda_2$ decays faster than the APT result for $\lambda_1$, a longer process duration $\tau$ is needed for such a decay to be achieved \cite{Morita2008,Rezakhani2010,Jansen2007,Venuti2018}.

The protocol $\lambda_3$ of Eq.~\eqref{eq:crossing_lambda3} has the same first derivatives at the end points as the linear protocol, resulting in the same APT scale of Eq.~\eqref{eq:APT1ExcessWork}.
However, $\lambda_3$ contrasts with $\lambda_1$ in its derivative at the time of crossing the QCP: $\lambda_3$ ``pauses'' the evolution at the crossing, which invalidates the LZ formula \eqref{eq:LZFormula} and, consequently, the KZM and LZF results of Eqs.~\eqref{eq:crossing_KZMExcessWork} and \eqref{eq:crossing_LZFExcessWork}.
This is shown in Fig.~\ref{fig:crossing_lambda13_APT}.
It is noteworthy that the pause effectively shortens the minimum duration necessary for APT to hold.
This corroborates the idea that pausing the variation of the field at a QCP improves the probability of finding the system in its ground state at the end of the process.
The same is true for nonunitary dynamics \cite{Marshall2019,Chen2020PRAppl,Passarelli2019}, but it should be noted that our result depends on knowing exactly where the QCP is \cite{Slutskii2019}, as is known in the TI chain.

\begin{figure*}

\subfloat[\label{fig:crossing_lambda14_LRTlin}]{\includegraphics[width=.5\textwidth]{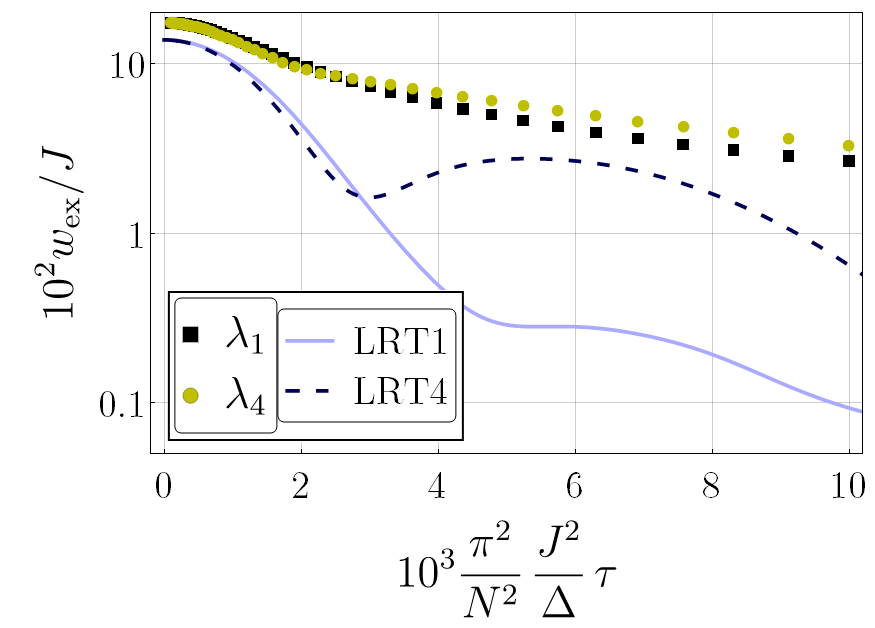}}
\subfloat[\label{fig:crossing_lambda14_LRT}]{\includegraphics[width=.5\textwidth]{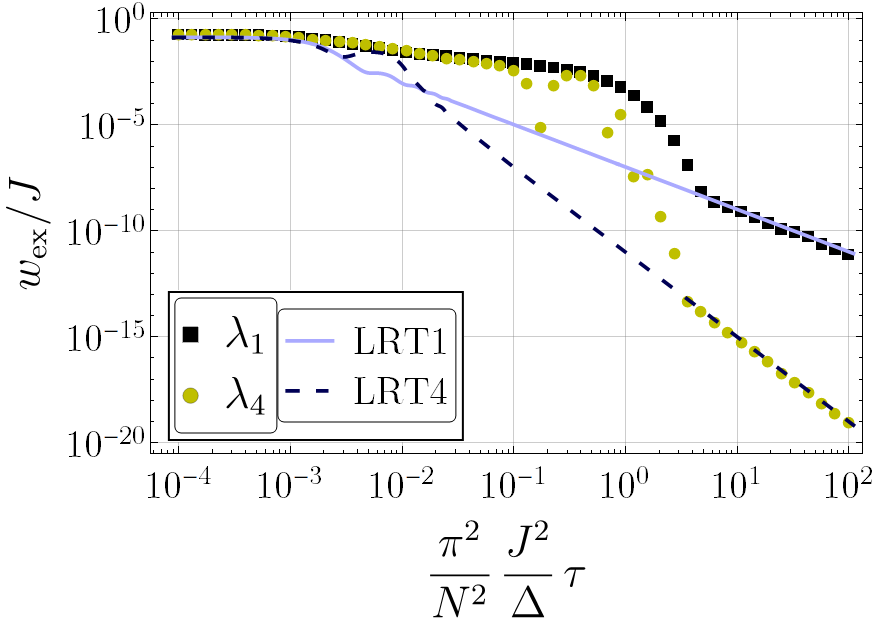}}

\caption{\label{fig:crossing_ExcessWorkPlots_LRT}
Plots of the excess work for the protocols of Eqs.~\eqref{eq:crossing_lambda1} and \eqref{eq:crossing_lambda4} as a function of process duration for $\Delta/J = 1$ and $N=100$.
The symbols represent the numerics for each protocol, while the lines represent LRT's results for the respective protocols.
(a) A linear-scale zoom of the region of very fast to sudden processes.
(b) The two protocols and LRT predictions for several decades, showing the full range of processes, from very fast to very slow.
}

\end{figure*}

Lastly, the protocol $\lambda_4$ of Eq.~\eqref{eq:crossing_lambda4} combines the previous two effects: It has null derivatives at the QCP \emph{and} at the end points.
Thus, unlike $\lambda_1$, it does not display KZM or LZF behavior, and also unlike $\lambda_1$, it decays as $\tau^{-4}$ in the region of slow processes.
This is shown in Fig.~\ref{fig:crossing_lambda14_APT}.

Thus, no matter the protocol, there is always an APT region (for finite chains).
Had we considered protocols with zero first \emph{and} second derivatives at the end points, then the excess work would scale as $\tau^{-6}$ for slow processes, and so on for higher derivatives \cite{Morita2008,Rezakhani2010}.
On the other hand, similar conclusions cannot be reached for the KZM and LZF results (at least not with a straightforward application of the KZM and LZF formulas).
Indeed, Fig.~\ref{fig:crossing_KZMregion} makes it very clear that protocols $\lambda_3$ and $\lambda_4$ do not display KZM behavior.
Driving too slowly through the QCP will inevitably suppress excitations and make the system reach the slow regime faster.

Regarding LRT, we first observe that, as $\tau\to 0$, the excess work converges to a value very close to the LRT prediction [see Sec.~\ref{sec:perturbationTheories_LRT} and Fig.~\ref{fig:crossing_lambda14_LRTlin}], no matter the protocol performed.
In addition, for very fast processes, LRT provides a solid lower bound for the excess work.

Also, we remark that the sudden-process case presents the highest values of excess work with respect to the switching time $\tau$.
This happens for two reasons: First, the relaxation function is an autocorrelation function \cite{Naze2021}, so the highest value occurs in the initial instant of time.
Second, the protocols are monotonic. Therefore 
\begin{equation}
    \frac{\Psi_{i}(0)\delta\lambda^2}{2} \ge \frac{1}{2} \int_{t_i}^{t_f} \int_{t_i}^{t_f} \Psi_i(t-t') \dot{\lambda}(t) \dot{\lambda}(t') dt' dt,
\end{equation}
as we observe in Fig.~\ref{fig:crossing_lambda14_LRT}.
We remark that this inequality is not always true for non-monotonic protocols \cite{bonanca2015}.

Additionally, one might think that the LRT results would fail in the TI chain if $\Delta/J \sim 1$.
However, that is not the case: Figure~\ref{fig:crossing_lambda14_LRT} shows that not only does LRT approximate well enough the sudden processes, but also it matches the numerics in slow processes for crossing the QCP with $\Delta/J = 1$.
As a matter of fact, LRT reproduces APT's result, correctly predicting the $\tau$ scales for slow processes.

The reason for this is the following: In the \hyperref[sec:APTandLRTagree]{Appendix}, we show the agreement between LRT and APT for simultaneously weak and slow processes.
This agreement rests in the fact that the weak limit ($\Delta \ll J$) of APT is obtained by replacing every instance of $\lambda(t)$ [but not $\dot{\lambda}(t)$] by $\lambda_i$ in APT's formulas, which makes sense when the system's Hamiltonian does not change considerably during the process.
Nevertheless, if $\lambda_f$ is not close to $\lambda_i$, but for some other reason $\epsilon_k(\lambda_f) \approx \epsilon_k(\lambda_i)$, then the approximation taken in the weak limit still applies, even without weak driving.
This is precisely what happens here: Since our protocols for crossing the QCP are symmetric with respect to it, the previously mentioned approximate equality between the initial and final dispersions holds true (at least for small values of $k$), and ultimately, LRT gives good results.
Notwithstanding, there is a tiny shift between data and LRT, a consequence of the imperfect approximation $\epsilon_k(\lambda_f) \approx \epsilon_k(\lambda_i)$ for nonsmall $k$.

\subsection{\label{sec:ComNum_stop}Stopping at the critical point}

\begin{figure*}

\subfloat[\label{fig:stopping_lambda12_APT}]{\includegraphics[width=.5\textwidth]{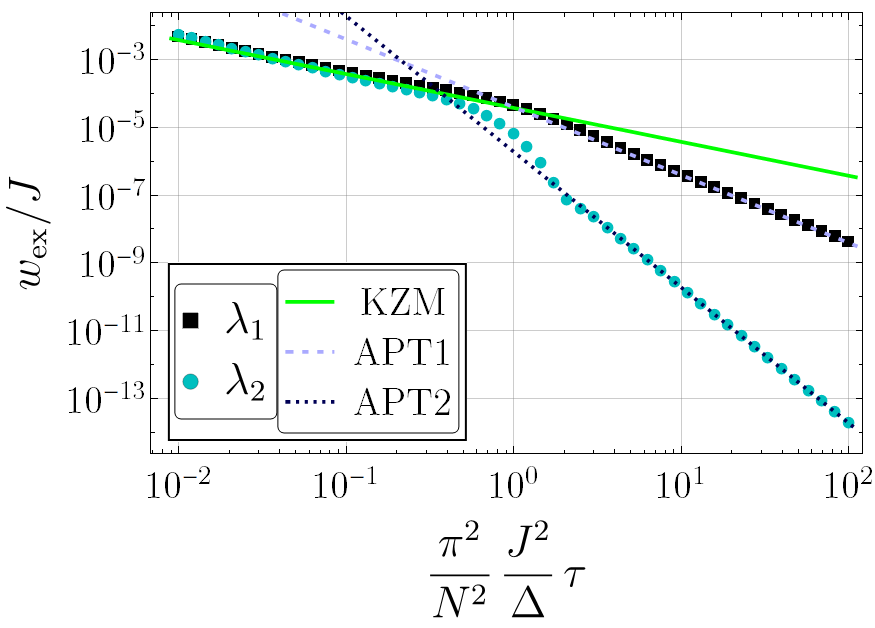}}
\subfloat[\label{fig:stopping_KZMregion}]{\includegraphics[width=.5\textwidth]{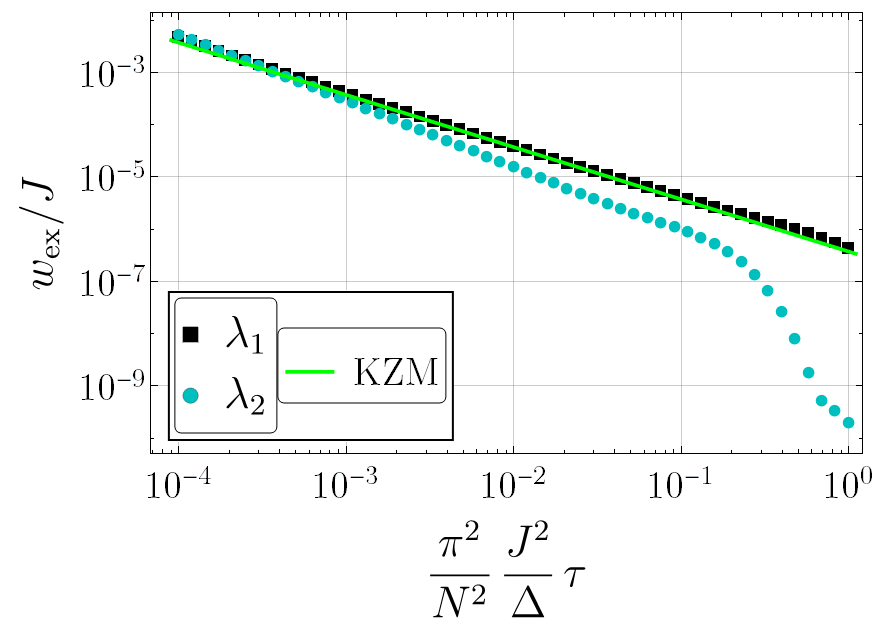}}

\caption{\label{fig:stopping_ExcessWorkPlots_APT}
Plot of the excess work for the protocols of Eqs.~\eqref{eq:stopping_lambda1} and \eqref{eq:stopping_lambda2} as a function of process duration for $\Delta/J = 1$.
The symbols represent the numerics for each protocol; the green line represents KZM of Eq.~\eqref{eq:stopping_KZMExcessWork}; the light blue, medium-dashed line represents first-order APT of Eq.~\eqref{eq:APT1ExcessWork}; and the dark blue, short-dashed line represents second-order APT of Eq.~\eqref{eq:APT2ExcessWork}.
(a) The two protocols and KZM and APT predictions for $N=100$.
(b) The two protocols and KZM for $N=1000$, where the agreement between the linear protocol $\lambda_1$ and KZM is better portrayed, while the nonlinear protocol $\lambda_2$ disagrees with KZM.
}

\end{figure*}

Since the condition $\dot{\lambda}(t_f) = 0$ for first-order APT to vanish is also the condition for the breakdown of the KZM expression in Eq.~\eqref{eq:stopping_KZMExcessWork}, the following two protocols will suffice for our discussion of the scenario of stopping at the critical point:
\begin{subequations}
\begin{align}
\label{eq:stopping_lambda1}
\lambda_1(t) & = \frac{t}{\tau}, \\
\label{eq:stopping_lambda2}
\lambda_2(t) & = - 8 \left( \frac{t}{\tau} \right)^5 - 20 \left( \frac{t}{\tau} \right)^4 - 18 \left( \frac{t}{\tau} \right)^3 - 7 \left( \frac{t}{\tau} \right)^2,
\end{align}
\end{subequations}
both for $t_i = -\tau$ and $t_f = 0$.
The protocol $\lambda_2$ of Eq.~\eqref{eq:stopping_lambda2} is the same as that of Eq.~\eqref{eq:crossing_lambda2}, but translated to fit the end points of this scenario, $\lambda(t_i) = -1$ and $\lambda(t_f) = 0$.
Therefore it has zero first derivatives at the end points, which breaks the KZM expression of Eq.~\eqref{eq:stopping_KZMExcessWork}, while also making its APT result be that of Eq.~\eqref{eq:APT2ExcessWork}.
This is shown in Fig.~\ref{fig:stopping_lambda12_APT}.
Here, protocols $\lambda_1$ and $\lambda_2$ contrast in every region of the plot.
Figure~\ref{fig:stopping_KZMregion} shows the intermediate range for a larger value of $N$, confirming that the nonlinear protocol indeed breaks KZM.

\begin{figure}[t]

\includegraphics[width=\columnwidth]{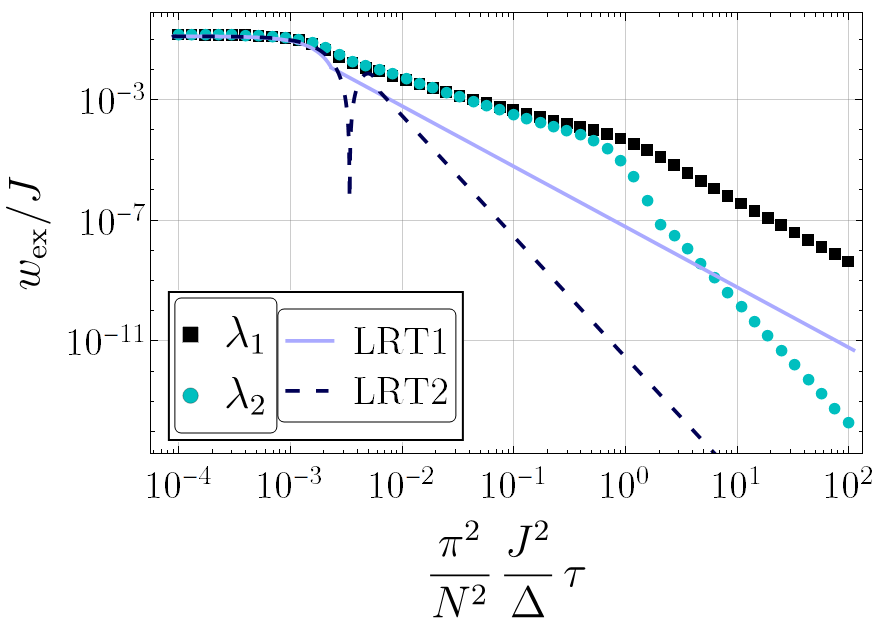}

\caption{\label{fig:stopping_lambda12_LRT}
Plot of the excess work for the protocols of Eqs.~\eqref{eq:stopping_lambda1} and \eqref{eq:stopping_lambda2} as a function of process duration for $\Delta/J = 1$ and $N=100$.
The symbols represent the numerics for each protocol, while the lines represent LRT's results for the respective protocols.
}

\end{figure}

This scenario shows how dependent LRT is on the symmetry of the protocols with respect to the QCP, or at least its long-$\tau$ behavior.
Figure~\ref{fig:stopping_lambda12_LRT} shows the case for stopping at the QCP, which is clearly an asymmetric evolution.
In this case, there is a big discrepancy between LRT and numerics, even though LRT still predicts the correct scale for the protocols considered.
The same observations that we have made for the sudden-process case in the previous section hold here.

\section{\label{sec:Conclusion}Concluding remarks}

In this paper, we have demonstrated that it is possible to significantly alter the behavior of the excitations of the transverse field Ising chain, as a function of driving duration $\tau$, with suitable choices for the time dependence of the external field.
In particular, we have shown that, with a slow enough passage through the quantum critical point of the system, one is able to prevent the appearance of Landau-Zener transitions, the source of universal Kibble-Zurek scaling in this system.
Ultimately, a well-placed pause in the variation of the field shortens the minimum driving duration required for adiabaticity to be attained, in the sense of adiabatic perturbation theory.
We have also seen that, while the power-law decay in $\tau$ for very slow processes can be made steeper by making the protocol vary smoothly at the beginning and end of the evolution, this comes at the price of delaying such decay. For almost sudden process, we observed that the excess work converges to its maximum value regardless of the protocol used, and that this maximum can be well estimated by linear response theory.

Our conclusions hold both for the scenario of crossing the critical point and for the scenario of stopping at the critical point.
However, the description of the excitations from linear response theory is specific for the first of these two scenarios, given the implied symmetry of the system's energy spectrum around the critical point. Nevertheless, when crossing the critical point, linear response theory surprisingly gives good lower bounds for extremely fast processes and captures the correct power-law decay for very slow processes even beyond its expected range of validity.

Our findings may have far-reaching ramifications for quantum annealing.
As has been noted previously \cite{Marshall2019,Chen2020PRAppl}, strategic pauses in annealing protocols may dramatically improve the performance.
In addition, we have highlighted that symmetric driving with respect to QCPs can further suppress excitations.
Hence further work may reveal universally applicable design guidelines of optimal driving strategies in adiabatic quantum computers.

\begin{acknowledgments}
A.S. and M.V.S.B. acknowledge E. Miranda for fruitful discussions.
A.S. and M.V.S.B. thank the National Council for Scientific and Technological Development (CNPq) for support under Grant No. 140549/2018-8 and FAEPEX (Fundo de Apoio ao Ensino, \`a Pesquisa e \`a Extens\~ao, Brazil) for support under Grant No. 2146-22.
P.N. and M.V.S.B. also acknowledge the financial support of CNPq (Grant No. 141018/2017-8) and FAPESP (Funda\c{c}\~ao de Amparo \`a Pesquisa do Estado de S\~ao Paulo, Brazil; Grants No. 2018/06365-4, No. 2018/21285-7, and No. 2020/02170-4).
B.G. acknowledges support from the National Science Center (NCN), Poland, under Project No. 2020/38/E/ST3/00269.
S.D. acknowledges support from the U.S. National Science Foundation under Grant No. DMR-2010127.
\end{acknowledgments}

\appendix*

\section{\label{sec:APTandLRTagree}Agreement between LRT and APT}

In this appendix, we prove that the expressions for the excess work from LRT and from APT are identical, for processes that are both weak and slow. The excess work, when calculated by means of LRT, is given by Eq.~\eqref{eq:LRTExcessWork} and we want to find the adiabatic limit of this expression, i.e., the leading order in $\tau^{-1}$ for $\tau \to \infty$. We can integrate the inner integral in Eq.~\eqref{eq:LRTExcessWork} by parts,
\begin{multline*}
\dot{\lambda}(t) \int \Psi_i(t-t') \dot{\lambda}(t') dt' = \dot{\lambda}(t) \dot{\lambda}(t') \int_{t}^{t'} \Psi_i(t-t'') dt'' \\
- \dot{\lambda}(t) \int \ddot{\lambda}(t') \int_{t}^{t'} \Psi_i(t-t'') dt'' dt',
\end{multline*}
and since $\dot{\lambda}(t) = O(\tau^{-1})$ and $\ddot{\lambda}(t) = O(\tau^{-2})$, the second term on the right-hand side is of a higher order than the first.
Therefore we can keep only the first term, and the same applies to the outer integration of Eq.~\eqref{eq:LRTExcessWork}.
Doing this carefully, the excess work reduces to
\begin{multline} \label{eq:LRTExcessWork_Adiabatic}
w_{\mathrm{ex}}^\mathrm{LRT}(\tau) \approx \frac{1}{2} \biggl[ \left( \dot{\lambda}^2(t_f) + \dot{\lambda}^2(t_i) \right)  \Upsilon_i(0)\\
- 2 \dot{\lambda}(t_f) \dot{\lambda}(t_i) \Upsilon_i(\tau) \biggr],
\end{multline}
where $\Psi_i$ and $\Upsilon_i$ are related by
\begin{equation}
\Psi_i(t) = - \ddot{\Upsilon}_i(t).
\end{equation}
For the TI chain, we can use the expression for $\Psi_i(t)$ in Eq.~\eqref{eq:RelaxationFunction} to arrive at
\begin{multline} \label{eq:TI_LRTExcessWork_Adiabatic}
w_{\mathrm{ex}}^\mathrm{LRT}(\tau) \approx \sum_{k>0} \frac{ \Delta^2 J_k^2 }{8N \epsilon^5_k(\lambda_i) } \biggl[ \dot{\lambda}^2(t_f) + \dot{\lambda}^2(t_i) \\
- 2 \dot{\lambda}(t_f) \dot{\lambda}(t_i) \cos\left( 2 \epsilon_k(\lambda_i) \tau \right) \biggr].
\end{multline}

On the other side, we have the excess work calculated through first-order APT, given in Eq.~\eqref{eq:APT1ExcessWork}, and we want to find its weak limit, i.e., the leading order in $\Delta$ for $\Delta \to 0$.
In that case, all we need to do is to replace $\lambda(t)$ by $\lambda_i$ in every quantity appearing in Eqs.~\eqref{eq:APT1ExcessWork}--\eqref{eq:APT2ExcessWork}, including the dispersion appearing inside the integral of Eq.~\eqref{eq:TI_DynamicPhase}.
We have
\begin{equation} \label{eq:TI_APTExcessWork_Weak}
w_{\mathrm{ex}}^\mathrm{APT1}(\tau) \approx \sum_{k>0} \frac{\Delta^2 J_k^2}{8N \epsilon_k^5(\lambda_i)} \left| \dot{\lambda}(t_f) - e^{-2i\epsilon_k(\lambda_i) \tau} \dot{\lambda}(t_i) \right|^2.
\end{equation}
Finally, expanding the squared absolute value in this last expression shows that Eqs.~\eqref{eq:TI_LRTExcessWork_Adiabatic} and \eqref{eq:TI_APTExcessWork_Weak} are indeed equal.
While shown here for the TI chain, the agreement between the theories is valid for any system, following the same arguments presented here.

\bibliography{bibliography}

\end{document}